# Threshold Logic with Current-Driven Magnetic Domain Walls

Xuan Hu[1], Brighton A. Hill[1], Felipe Garcia-Sanchez[2], and Joseph S. Friedman[1]

[1]Electrical & Computer Engineering, The University of Texas at Dallas, Richardson, TX, United States
[2]Universidad de Salamanca, Salamanca, Spain

*Abstract*--The recent demonstration of current-driven magnetic domain wall logic [Z. Luo *et al.*, *Nature* 579:214] was based on a three-input logic gate that was identified as a reconfigurable NAND/NOR function. We reinterpret this logic gate as a minority gate within the context of threshold logic, enabling a domain wall threshold logic paradigm in which the device count can be reduced by 80%. Furthermore, by extending the logic gate to more than three inputs of non-equal weight, an 87% reduction in device count can be achieved.

Threshold logic [1] efficiently implements complex logical functions within a single gate, but its widespread adoption has been impeded by insufficient nonlinearity in conventional transistors. The recent experimental demonstration of cascaded logic gates with current-driven magnetic domain walls [2] operates based on the nonlinearity intrinsic to anisotropic systems, thus naturally enabling threshold logic gates. However, these logic gates are described in [2] as reconfigurable NAND and NOR gates, rather than as three-input minority threshold logic gates.

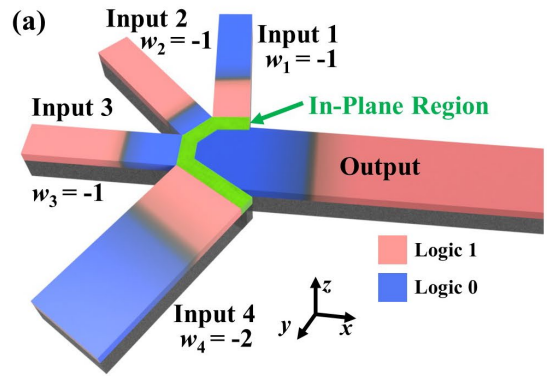

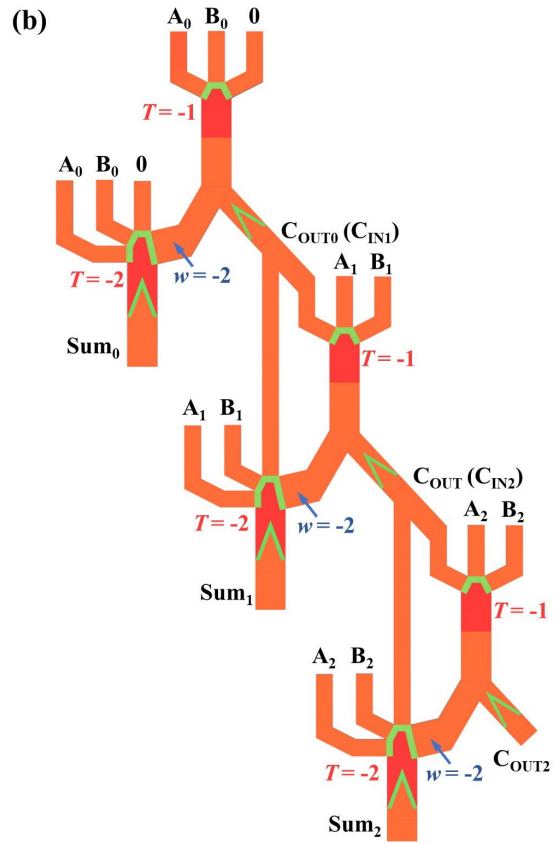

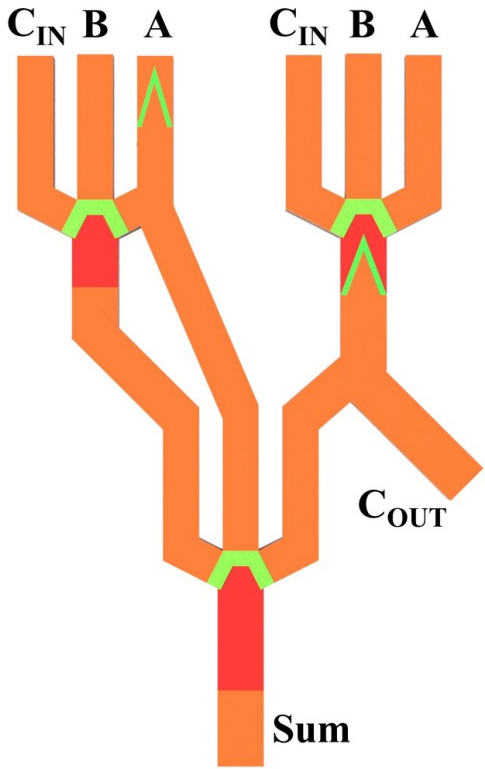

Fig. 1. One-bit full adder implemented with only three DWTL minority gates.

Fig. 2. (a) Four-input DWTL gate in which three inputs have weight $w$=-1, and one input has $w$=-2. (b) Three-bit full adder with only two DWTL gates per bit (six total).

We therefore propose a novel domain wall threshold logic (DWTL) paradigm that enables circuits that are simpler and more efficient than can be achieved with the reconfigurable gates of [2]. Similar to [2], magnetic domains magnetized in the $+z$ ($-z$) direction represent logic '1' ('0'), and each in-plane region causes magnetization reversal due to chiral coupling resulting from the interfacial DMI. As demonstrated in [2], DWTL gates produce an output domain that is the inversion of the majority of the inputs. That is, this logic gate performs the minority threshold function, which is significantly more logically expressive than NAND or NOR. By leveraging this minority gate behavior, the one-bit full adder of Fig. 1 can be realized with three logic gates, 80% fewer than the 15 required in [2].

We also propose DWTL functions with logical expressivity increased further by using more than three inputs, and with inputs that have non-uniform impacts on the output magnetization. As shown in Fig. 2a, a four-input DWTL gate can be realized in which one input has twice the width of the other three; therefore, this wider input will have roughly twice the impact on the output magnetization. DWTL gates with greater logical expressivity further reduce the number devices required to perform logical operations, such as in the three-bit DWTL full adder of Fig. 2b that requires only two gates per bit (87% reduction).